# Medical Knowledge Integration Into Reinforcement Learning Algorithms for Dynamic Treatment Regimes

Sophia Yazzourh[1] , Nicolas Savy[1] ,
Philippe Saint-Pierre[1] and Michael R. Kosorok[2]

[1]Institut de Mathématiques de Toulouse; UMR5219 - Université de Toulouse, Toulouse, France
[2]Department of Biostatistics, University of North Carolina at Chapel Hill, Chapel Hill, North Carolina, USA
Corresponding to: Sophia Yazzourh, Institut de Mathématiques de Toulouse; UMR5219 - Université de Toulouse, Toulouse, France. Email: sophia.yazzourh@math.univ-toulouse.fr

## Summary

The goal of precision medicine is to provide individualised treatment at each stage of chronic diseases, a concept formalised by dynamic treatment regimes (DTR). These regimes adapt treatment strategies based on decision rules learned from clinical data to enhance therapeutic effectiveness. Reinforcement learning (RL) algorithms allow to determine these decision rules conditioned by individual patient data and their medical history. The integration of medical expertise into these models makes possible to increase confidence in treatment recommendations and facilitate the adoption of this approach by healthcare professionals and patients. In this work, we examine the mathematical foundations of RL, contextualise its application in the field of DTR, and present an overview of methods to improve its effectiveness by integrating medical expertise.

## 1 Introduction

Modern medicine, with its remarkable advancements in care, drugs, and treatments, now seeks to enhance its ability to deliver personalised treatments for each individual patient. The paradigm of precision medicine (Kosorok & Laber, 2019) initiates a profound consideration of this question. Precision medicine aims to optimise the quality of healthcare by tailoring the medical approach to match the specific and continually changing health condition of every individual patient. The heterogeneity among patients' populations and sub-populations leads to distinct reactions and, consequently, necessitates different treatment approaches. Initially, this research domain introduced statistical models (Chakraborty & Murphy, 2014; Kosorok & Laber, 2019; Kosorok & Moodie, 2015) aimed at facilitating decision-making support. Naturally, with the advent of data

This work was supported by the ANR LabEx CIMI (grant ANR-11-LABX-0040) within the French State Programme 'Investissements d'Avenir'.

storage and the computational power, machine learning methods (Coronato *et al.*, 2020; Yu *et al.*, 2021) have also begun to be applied to address this issue.

In this context, one of the growing interests of modern medicine is to adapt prescribed treatments based on individual data, unique characteristics and particular medical history of the patient. Precision medicine seeks to put the patient's own information at the centre in order to improve their health. The motto behind is 'The right treatment for the right patient (at the right time)'. In a 2015 State of the Union address, President Obama announced a Precision Medicine Initiative to revolutionise how we improve health, research, and treat disease. The initiative defines precision medicine as 'an emerging approach for disease treatment and prevention that takes into account individual variability in genes, environment, and lifestyle for each person' (Terry, 2015). In technical terms, adaptive treatment strategies or dynamic treatment regimes (DTR) formalise the objective of enhancing the care pathway for patients by proposing an optimal and personalised treatment sequence. They aim to establish a decision rule at each stage of the care process. It conditions the treatment based on responses to previous prescriptions and medical history (Chakraborty & Murphy, 2014; Laber *et al.*, 2014). The goal is to optimise the patient's long-term positive response to the sequence of treatment decisions while tailoring the treatment to their own medical information (Kosorok & Moodie, 2015).

In the past decades, machine learning has emerged as a solution to large-scale and high-complexity problems. When it comes to decision support, particularly in sequential scenarios, reinforcement learning (RL) (Sutton & Barto, 2018) offers the most effective solution. These methods excel in adapting to changing conditions and optimising decisions over a series of steps, making them especially valuable in dynamic decision-making processes. The concept revolves around identifying a decision rule, referred to as policy, which is designed to optimise a long-term objective. This policy is crafted in order to make decisions over time that lead to the greatest cumulative benefit or outcome.

RL methods are thus an appealing candidate for precision medicine and have been extensively studied as a potential tool for guiding medical decisions towards personalised treatment. First, the application of these methods to DTR is facilitated by modelling the underlying decision problem using a so-called decision process (DP), as detailed in Section 2. It is straightforward to express and establish connections between medical elements and its mathematical components. Second, the primary aim of RL is to identify this decision rule. In this context, there is a desire to establish this rule while maximising long-term cumulative gains. In medicine, the effects of treatments and side effects are not immediate but can take several stages to manifest. The way the policy is constructed is a significant asset for precision medicine. Third, RL models have the capacity to simultaneously consider the extensive patient covariates data and address multi-stage decision problems. The scope of RL applications in precision medicine is in recent thematic reviews of major interest: a non-technical survey offering illustrations of RL applications in public health is proposed in Weltz *et al.* (2022). More specifically, RL applications in the context of mobile health are presented in Deliu *et al.* (2024). Two more technical reviews describe the methods for determining medical decision rules using off-policy RL approach (Uehara *et al.*, 2022), or more specifically with the use of Q-learning (Clifton & Laber, 2020) and their empirical comparison with other estimation methods (Li *et al.*, 2023).

While RL offers promising algorithms for sequential decision-making in healthcare, as detailed in the Data S1, relying on a machine learning algorithm may create apprehension among all stakeholders in the process. This hesitation can originate from both the patient and the physician sides. In order to be operational in a clinical context, several points must be improved such as safer, more interpretative and efficient medical decision making (Eckardt *et al.*, 2021). One approach to enhance the application of RL in healthcare is the integration of expertise or human knowledge into the models. The concept is to create a partnership between both machine learning capabilities and domain experts (Holzinger, 2016; Maadi *et al.*, 2021). This 'collaboration' would not only improve confidence in RL models and the recommendations they provide (Love *et al.*, 2023) but also facilitate the utilisation of this technology by healthcare professionals and patients within a clinical setting (Holzinger *et al.*, 2019). This merging of machine learning and human expertise yields to improved results compared with RL in isolation or expert decisions alone (Arzate Cruz & Igarashi, 2020; Li *et al.*, 2019). From a technical point, involving experts or medical knowledge also reduces the learning time, allowing for quicker adaptation and enhancement of the methods, ultimately leading to more effective and patient-centred healthcare solutions.

The objective of this paper is to provide a comprehensive overview of RL applied to the optimisation of treatment sequences. By facilitating an entry into this field for those interested in its practical application in precision medicine, we illustrate its mathematical framework and provide contextualisation. This overview aims to help navigate the array of available algorithms. Additionally, we explore the integration of medical knowledge into RL models, highlighting considerations that could facilitate their clinical integration and application. We introduce these issues, offering initial questions and showcasing opportunities for further research in this area.

To achieve this objective, we structure the paper as follows. In Section 2, we delve into the mathematical foundations of RL approaches, specifically exploring decision processes and introducing key concepts and specific terms of the domain: policy, rewards, and value function. In Section 3, we contextualise this study within the realm of DTR, offering a more detailed explanation of how RL and the concept of precision medicine are intricately connected. We also explain the properties and classification of RL algorithms within our medical context. In Section 4, we provide an overview of methods to enhance RL in the medical context by integrating expert knowledge. Various methods are presented and discussed. The paper ends by a concluding Section 5.

## 2 Theoretical Foundations of Reinforcement Learning

This section aims to outline the mathematical framework of RL applied in the DTR field. Typically, RL is explained in the context of a Markov decision process (MDP) and its evolution into a partially observable Markov decision process (POMDP). However, in this context, a return is made to a decision-making framework without the inclusion of Markov assumptions, which is referred to as a decision process. Subsequently, fundamental concepts are introduced: policy, value function, and the notion of optimality.

### 2.1 Decision Process

#### 2.1.1 General statement

The modelling context revolves around the realm of decision-making. A foundation proposed is DP, which acts as the initial framework for DTR. It represents a dynamic system which evolves through time $t \in \mathbb{T}$. This system navigates within the state space $\mathbb{S}$ by executing actions within the realm of possibilities defined by the space of actions $\mathbb{A}$. The collection of non-empty measurable subsets of $\mathbb{A}$, denoted as $\{\mathbb{A}(s)|s \in \mathbb{S}\}$, represents the feasible actions that can be undertaken when the system finds itself in a specific state $s \in \mathbb{S}$.

**Definition 2.1.** *Decision process. A decision process* $(S, A, \{\mathbb{A}(s)|s \in \mathbb{S}\}, v)$ *on* $\mathbb{T}$ *includes*

- a family $S$ of $\mathbb{S}$-valued random variables $\{S_t, t \in \mathbb{T}\}$, where $\mathbb{S}$ is called the state space.
- a family $A$ of $\mathbb{A}$-valued random variables $\{A_t, t \in \mathbb{T}\}$, $\mathbb{A}$ is called space of actions.

- a family $\{\mathbb{A}(s)|s \in \mathbb{S}\}$ of non empty measurable subsets of $\mathbb{A}$, the set of realisable actions when the system is in the state $s \in \mathbb{S}$. The requirement is for $\mathbb{K} = \{(s, a)|s \in \mathbb{S}, a \in \mathbb{A}(s)\}$ to be a measurable subset of $\mathbb{S} \times \mathbb{A}$.
- a distribution $v$ on $\mathbb{S}$.

**Remark 2.1.** *DP is initially characterised for Borel spaces $\mathbb{S}$ and $\mathbb{A}$. However, in most practical applications, these spaces are typically finite-dimensional, a context that will be considered for the remainder of the article. This theoretical framework is inspired by the works of Hernández-Lerma & Lasserre (2012) and Nivot (2016), particularly in the discrete setting. However, it is worth noting that these references focus on a Markovian framework, which will be discussed in details later in the article.*

**Remark 2.2.** *$S_t$ represents the state of the system at time $t$. In general, the state space is denoted by $\mathbb{S}$. When the state space has a linear structure, $\mathbb{S}$ can be considered to be $\mathbb{R}^d$, where $d$ is the dimension of the Euclidean space, corresponding to the number of components considered. In this setting, $S_t$ is modeled as a vector of covariates observed at time t.*

**Remark 2.3.** *In full generalities, $\mathbb{T}$ will be taken as continuous or discrete but for a sake of readability $\mathbb{T}$ will be a discrete space denoted by $\mathbb{T} = \{0 = t_0, t_1, \ldots, t_n, \ldots, \tau\}$, with $\tau$ representing either a finite ($\tau = t_N < \infty$) or infinite ($\tau = \infty$) value. For the sake of simplicity, the variables $X_{t_n}$ will be indicated as $X_n$ and $X_\tau$ as $X_\infty$ in infinite horizon setting.*

**Definition 2.2.** *For any $n \in \mathbb{N}$, an admissible history at time $n$ is a vector which contains the states travelled by the system together with the actions taken up to time $n$. Let $\mathbb{K} = \{(s, a)|s \in \mathbb{S}, a \in \mathbb{A}(s)\}$ denotes the subset of state-action pairs. The set of admissible histories at time n is denoted:*

$$\mathbb{H}_0 = \mathbb{S} \qquad \mathbb{H}_n = \mathbb{K}^{n-1} \times \mathbb{S}$$

An element $h_n \in \mathbb{H}_n$ writes $(s_0, a_0, \ldots, s_{n-1}, a_{n-1}, s_n)$ where for all $0 \leq j \leq n-1$, $(s_j, a_j) \in \mathbb{K}$.

The point of main importance to deal with a decision process is to exhibit the probability to reach state $s_{n+1}$ at time $n+1$ given the history up to time $n$ and the decision taken at time $n$ this expresses as

$$\mathbb{P}_v[S_{n+1} = s_{n+1}|H_n = h_n, A_n = a_n]. \tag{1}$$

where $\mathbb{P}_v$ represents the transition probability conditioned on the initial distribution $v$.

In practice the computation of these probabilities requires significant computational resources because of the increasing length of the vector $h_n$ as $n$ increases. Rapidly working directly with such variable is intractable (usually when $n \geq 4$).

### 2.1.2 Markov decision process

To overpass this difficulty the Markov assumption is of particular interest. It consists in simplifying the dependence on the past by considering that all the necessary information for is contained in the current state.

**Definition 2.3.** *Markov decision process. A Markov decision process on $\mathbb{T}$ is a decision process $(\mathbb{S}, \mathbb{A}, \{\mathbb{A}(s)|s \in \mathbb{S}\}, v)$ satisfying*

$$\mathbb{P}_v[S_{n+1} = s_{n+1} | H_n = h_n, A_n = a_n] = \mathbb{P}_v[S_{n+1} = s_{n+1} | S_n = s_n, A_n = a_n]. \quad (2)$$

A MDP is thus governed by a family of transitions probabilities

$$P_{a_n}(s_n, s_{n+1}) = \mathbb{P}_v[S_{n+1} = s_{n+1} | S_n = s_n, A_n = a_n].$$

which is the probability that action $a_n$ in state $s_n$ at time $t_n \in \mathbb{T}$ leads to state $s_{n+1}$ at time $t_{n+1}$.

The most traditional RL framework is MDP (Bellman, 1957; Garcia & Rachelson, 2013). The majority of optimising application complete their decision models with the memory-less Markov assumption.

**Remark 2.4.** *It is also common to see Definition 2.1 extended to the framework of a Markov Decision Process supplemented by a transition kernel denoted $\mathbb{P}_v$.*

**Remark 2.5.** *Behind MDP modelling lies a strong assumption that all the information necessary for decision-making is fully observed. In practice, however, the state space is often noisy or incomplete. Unobserved confounders represent one of the greatest challenges for real-world medical applications, partly due to the way clinical data are collected.*

One approach to relaxing this assumption is to use the Partially Observable Markov Decision Process (POMDP) model, introduced in Monahan (1982). POMDPs can be seen as a generalisation of MDPs, building upon the same formal framework but allowing for partial observability. The major difference lies in the structure of the state space: while classical DP and MDP frameworks assume that all relevant variables are directly observed, POMDPs explicitly distinguish between observed and unobserved components. Mathematically, a POMDP is defined similarly to an MDP, except that the state $S_n$ at time $n$ consists of a pair $(S_n^{obs}, S_n^{unobs})$ taking values in $\mathbb{S}^{obs} \times \mathbb{S}^{unobs}$, where $S_n^{obs}$ is observed and $S_n^{unobs}$ remains hidden.

A second approach is provided by the Confounded MDP framework (Fu *et al.*, 2022; Stensrud *et al.*, 2024), which extends the standard MDP formalism to explicitly account for unobserved confounders. In classical MDPs, transition dynamics are modelled as depending only on the observed state and action. In contrast, Confounded MDPs assume that transitions are also influenced by latent variables. Formally, a Confounded MDP is defined as an MDP except that the transition probability is given by $\mathbb{P}(S_{n+1}|S_n, A_n, U_n)$, where $U_n$ represents unobserved confounders that are not part of the observed state space.

In both approaches, medical expertise plays a crucial role. Experts can help identify clinically relevant variables that should be measured to reduce the risk of unobserved confounding and provide insights into potential sources of hidden biases. For example, if information on treatment decisions made by clinicians is available, it can be incorporated as an additional covariate to supplement the information used for RL. In some cases, instrumental variables based on clinical practice patterns can also be leveraged to further mitigate the impact of unobserved confounding. Incorporating domain knowledge at both the modelling and data collection stages is thus essential to enhance the validity of the decision-making framework.

*2.2 Policy*

The crucial concept in addressing dynamic programming is the notion of a policy, which is formalised as follows:

A policy is a sequence $\pi = (\pi_n)_{n \in \mathbb{N}}$ of conditional distributions from $\mathbb{A}$ given $\mathbb{H}_n$ defined, for any $\mathcal{A} \in \mathcal{B}(\mathbb{A})$ and all $h_n \in \mathbb{H}_n$, by

$$\pi_n(\mathcal{A}, h_n) = \mathbb{P}_\nu[A_n \in \mathcal{A} | H_n = h_n],$$

satisfying for all $n \in \mathbb{N}$, all $h_n \in \mathbb{H}_n$

$$\pi_n(\mathbb{A}(s_n), h_n) = 1,$$

and for all $n \in \mathbb{N}$, all $h_n \in \mathbb{H}_n$ and all $a_n \in \mathbb{A}(s_n)$

$$\pi_n(a_n, h_n) > 0.$$

Decision-making is selecting an option based on environmental information. A policy represents a plan that establishes a sequence of actions. This strategy can be tailored to align with a specified objective. As a result, the focus will be on deriving the strategy that optimises this objective. A policy $\pi_n$ is a strategy that suggests, for every possible states $s_n \in \mathbb{S}$, an action $a_n \in \mathbb{A}(s_n)$ taking to account the history $h_n \in \mathbb{H}_n$ of the system.

**Theorem 2.1.** *(Hernández-Lerma & Lasserre, 2012; Nivot, 2016) Given a policy $\pi$ and the initial distribution $\nu$, there is a unique probability $\mathbb{P}_\nu^\pi$ such that, for all $\mathcal{B} \in \mathcal{B}(\mathbb{S})$, the Borel algebra of $\mathbb{S}$, and $\mathcal{A} \in \mathcal{B}(\mathbb{A})$, the Borel algebra of $\mathbb{A}$:*

$$\mathbb{P}_\nu^\pi[S_0 \in \mathcal{B}] = \nu(\mathcal{B}),$$

$$\mathbb{P}_\nu^\pi[A_n \in \mathcal{A} | H_n = h_n] = \pi_n(\mathcal{A}, h_n)$$

In the following, $\mathbb{E}_\nu^\pi$ denotes the expectation associated with the probability $\mathbb{P}_\nu^\pi$ for an arbitrary policy $\pi$ and an initial distribution $\nu$.

The following result is of major practical importance and expresses the likelihood to observe a trajectory $h_n$ by means of the DP. Theorem 2.1 follows directly from the Ionescu–Tulcea theorem. This is a well-known result, and detailed proofs can be found in Hernández-Lerma & Lasserre (2012) and Nivot (2016).

**Theorem 2.2.** *Given $(S, A, \{\mathbb{A}(s) | s \in \mathbb{S}\}, \nu)$ a decision process on $\mathbb{T}$ and $\pi$ a policy, we have for all $n \in \mathbb{N}^*$ and all $h_n \in \mathbb{H}_n$*

$$\mathbb{P}_\nu^\pi[H_n = h_n] = \nu(s_0) \prod_{j=1}^n \mathbb{P}_\nu\big[S_j = s_j | A_{j-1} = a_{j-1}, H_{j-1} = h_{j-1}\big] \pi(a_{j-1}, h_{j-1})$$

In the framework of MDP, to follow the same lines as in the proof of Theorem 2.2, an additional assumption on the policy is needed yielding to the concept of Markov policy:

**Definition 2.4.** *Markovian policy. (Hernández-Lerma & Lasserre, 2012; Nivot, 2016) A Markovian policy $\pi = (\pi_n)_{n \in \mathbb{N}}$ is a policy satisfying for all $n \in \mathbb{N}$, all $\mathcal{A} \in \mathcal{B}(\mathbb{A})$ and all $h_n \in \mathbb{H}_n$*

$$\mathbb{P}_\nu[A_n \in \mathcal{A} | H_n = h_n] = \mathbb{P}_\nu[A_n \in \mathcal{A} | S_n = s_n] = \pi_n(\mathcal{A}, s_n).$$

## 2.3 Rewards, Valuation and Optimisation of Policies

### 2.3.1 Rewards

As discussed in the Introduction, the aim of DP modelling is to find optimal policies associated to an objective. To do so, a criterion of optimality has to be introduced. This criterion is usually built by means of rewards functions which provides a temporal judgement of the desirability of a state-action pair and are formalised as follows:

**Definition 2.5.** *Reward is defined as a family of bounded $\mathbb{R}$-valued random variables $\{R_n, n \in \mathbb{N}\}$. For the sake of simplicity, let us denote for a given $n \in \mathbb{N}$, for all $h_n \in \mathbb{H}_n$, all $a_n \in \mathbb{A}$ and all $s_{n+1} \in \mathbb{S}$*

$$\mathcal{R}_{n+1}(h_n, a_n, s_{n+1}) = \mathbb{E}_v^\pi[R_{n+1} | H_n = h_n, A_n = a_n, S_{n+1} = s_{n+1}]$$

where $\mathcal{R}_{n+1}$ is called the immediate reward function.

**Remark 2.6.** *Although reward functions are often included in the definition of a decision process, it can be useful to treat them separately. In this perspective, the decision process describes the dynamics of the environment and the information available to the agent, independently of any specific objective. The reward function is then introduced to define a goal, guiding the construction of a policy. This separation is particularly relevant in offline settings where the reward can be inferred from observed behaviour, as in inverse RL, discussed in Section 4.2.*

### 2.3.2 Valuation of policies and value-functions

State-value functions and state-action values functions are respectively known as V-function and Q-functions. These two concepts provide quantitative measures for evaluating policies, making meaningful policies comparisons and defining optimal policies. These value-functions serve as qualitative evaluations for guiding strategic adaptations.

State-value functions allow to answer to: 'How good is to be in state *s* after following the policy $\pi$?' while action-value functions allow to answer to: 'How good it is to have done the action *a* following policy $\pi$ knowing that they were in state *s*?'. The key point is that the evaluation is not assessing step-by-step performance, but is based on the cumulative reward over time. In such a way, value functions focus on a long-term objective.

**Definition 2.6.** *Given $\gamma < 1$ a discount parameter, the stage n long term discounted reward function is defined for all $n \in \mathbb{N}$, by*

$$G_n = \sum_{j=n+1}^{\infty} \gamma^{j-n-1} R_j$$

**Definition 2.7.** *Value functions (Chakraborty & Murphy, 2014; Schulte et al., 2014). Given $(S, A, \{\mathbb{A}(s)|s \in \mathbb{S}\}, v)$ a decision process on $\mathbb{T}$, $\{R_n, n \in \mathbb{N}\}$ a family of rewards, $\pi$ a policy and $\gamma < 1$ a discount parameter.*

- The stage *n* state-value function (V-function) for a history $h_n$ is the total expected future rewards from stage *n* given by

$$V_n^\pi(h_n) = \mathbb{E}_v^\pi[G_n | H_n = h_n].$$

- The stage *n* action-value function (Q-function) is the total expected future rewards starting from a history $h_n$, taking action $a_n$ is given by

$$Q_n^\pi(h_n, a_n) = \mathbb{E}_v^\pi[G_n | H_n = h_n, A_n = a_n].$$

The crucial aspect to observe in these definitions is that, instead of a step-by-step evaluation, the approach aims to assess a long-term objective. The goal is to evaluate the cumulative reward over time. As a consequence of a decision, after each time step $t_n$, an immediate reward $R_n$ is received which is the most distinctive feature of RL. The value functions represent the total expected future reward starting at a particular state $s_0$ and thereafter choosing actions according to the policy $\pi$.

**Remark 2.7.** *The discount factor $\gamma$ introduced in the definition of the long-term reward at each step n aims to strike a thoughtful balance between immediate rewards and long-term rewards. It allows for a balancing between striving for the highest cumulative reward and the aim to reach substantial benefits within a reasonable time (Coronato et al., 2020). This is also a mathematical trick to make the sum converge.*

**Remark 2.8.** *In the finite horizon case $\tau = t_N$, the values functions can be defined in a similar way by considering*

$$G_n = \sum_{j=n+1}^{N} \gamma^{j-n-1} R_j$$

Notice that in this framework, the introduction of a discount parameter is not needed and is usually fixed to 1 from the definitions.

**Remark 2.9.** *To consider evaluation in infinite horizon, we have considered processes in infinite horizon and to do so, the Markov assumptions on the decision process and on the policy are necessary. The discount factor is now mandatory to ensure the convergence of the long term discounted reward. The value functions can be defined in the same way by considering expectations conditioned on S:*

$$V_n^\pi(s_n) = \mathbb{E}_\nu^\pi[G_n | S_n = s_n].$$
$$Q_n^\pi(s_n, a_n) = \mathbb{E}_\nu^\pi[G_n | S_n = s_n, A_n = a_n].$$

The following proposition highlights the link between V-functions and Q-functions.

**Proposition 2.1.** *(Kosorok & Moodie, 2015; Schulte et al., 2014; Sutton & Barto, 2018) For all $n \in \mathbb{N}$, all $h_n \in \mathbb{H}_n$ and $a_n \in \mathbb{A}$, we have*

$$V_n^\pi(h_n) = \sum_{a_n \in \mathbb{A}(s_n)} Q_n^\pi(h_n, a_n) \pi_n(h_n, a_n) \tag{3}$$

$$Q_n^\pi(h_n, a_n) = \sum_{s_{n+1} \in \mathbb{S}} \left( \mathcal{R}_{n+1}(h_n, a_n, s_{n+1}) + \gamma V_{n+1}^\pi((h_n, a_n, s_{n+1})) \right) \tag{4}$$
$$\times \mathbb{P}_\nu^\pi[S_{n+1} = s_{n+1} | H_n = h_n, A_n = a_n].$$

The remaining issue consists in the computation of the value functions. To do so, the result of major importance is the recursive form of the value functions which states that the value functions can be decomposed into immediate reward plus discounted value of successor state.

**Theorem 2.3.** *Recursive form for value functions. (Chakraborty & Murphy, 2014; Zhao et al., 2015) For all $n \in \mathbb{N}$, all $h_n \in \mathbb{H}_n$ and $a_n \in \mathbb{A}$, we have*

$$V_n^\pi(h_n) = \sum_{s_{n+1} \in \mathbb{S}} \sum_{a_n \in \mathbb{A}(s_n)} \left(\mathcal{R}_{n+1}(h_n, a_n, s_{n+1}) + \gamma V_{n+1}^\pi(h_{n+1})\right) \quad (5)$$
$$\times \mathbb{P}_v[S_{n+1} = s_{n+1} | H_n = h_n, A_n = a_n]\pi(a_n, h_n)$$

$$Q_n^\pi(h_n, a_n) = \sum_{s_{n+1} \in \mathbb{S}} \left(\mathcal{R}_{n+1}(h_n, a_n, s_{n+1})\right. \quad (6)$$
$$+ \gamma \sum_{a_{n+1} \in \mathbb{A}(s_{n+1})} Q_{n+1}^\pi(h_{n+1}, a_{n+1})\pi(h_{n+1}, a_{n+1}))$$
$$\times \mathbb{P}_v[S_{n+1} = s_{n+1} | H_n = h_n, A_n = a_n]$$

Equations 5 and 7 are known as Bellman's equation. A policy being fixed, the Bellman equation can be solved, therefore making it possible to determine the values of the value functions and thus the values of Q-function. Indeed, in the case where the number of steps is finite, the Bellman equation actually hides a linear system of $N$ equations to $N$ unknowns, where $N$ is final finite number of steps considered. It can therefore be solved, once translated into a matrix equation, by a technique such as the Gaussian elimination.

### 2.3.3 Optimisation of the policies

The key concern of the RL problem is to determine an optimal policy, denoted as $\pi^*$, which represents the optimal strategy for maximising our long-term reward function. In other words, it is about finding the best way to make decisions in an environment to obtain the highest long-term rewards. The search for an optimal policy is based on the Bellman optimality principle developed below.

**Definition 2.8.** *An optimal state-value function $(V_n^*)$ is defined for all $n \in \mathbb{N}$, all $h_n \in \mathbb{H}_n$ as the maximum value functions over all policies*

$$V_n^*(h_n) = \max_\pi V_n^\pi(h_n)$$

An optimal action-value function $(Q_n^*)$ is defined for all $n \in \mathbb{N}$, all $h_n \in \mathbb{H}_n$ and $a_n \in \mathbb{A}$, as the maximum action-value functions over all policies

$$Q_n^*(h_n, a_n) = \max_\pi Q_n^\pi(h_n, a_n)$$

**Definition 2.9.** *Consider the partial ordering over policies defined by*

$$\pi' \geq \pi \text{ if and only if, for all } n \in \mathbb{N}, \text{ all } h_n \in \mathbb{H}_n, \ V_n^{\pi'}(h_n) \geq V_n^\pi(h_n).$$

This partial ordering allows to define optimal policies in the following way:

**Proposition 2.2.** *There exists a policy $\pi^*$ that is better than or equal to all other policies, i.e. $\pi^* \geq \pi$ for all $\pi$. Such a policy is called an optimal policy.*

**Theorem 2.4.** *All optimal policies achieve an optimal value functions and an optimal action-value functions, for all $n \in \mathbb{N}$, all $h_n \in \mathbb{H}_n$ and $a_n \in \mathbb{A}$,*

$$V_n^{\pi^*}(h_n) = V_n^*(h_n) \quad \text{and} \quad Q_n^{\pi^*}(h_n, a_n) = Q_n^*(h_n, a_n).$$

**Theorem 2.5.** *Bellman optimality equations for $Q_n^*$. For all $n \in \mathbb{N}$, all $h_n \in \mathbb{H}_n$ and $a_n \in \mathbb{A}$, we have*

$$Q_n^*(h_n, a_n) = \sum_{s_{n+1} \in \mathbb{S}} \left( \mathcal{R}_{n+1}(h_n, a_n, s_{n+1}) + \gamma \max_{a \in \mathbb{A}(s_{n+1})} Q_{n+1}^*(h_{n+1}, a) \right) \quad (7)$$
$$\times \mathbb{P}_v[S_{n+1} = s_{n+1} | H_n = h_n, A_n = a_n]$$

As a consequence of the Bellman Optimality Equation, we can claim that an optimal policy can be found by maximising over $Q_n^{\pi^*}(s, a)$ for all $n \in \mathbb{N}$ and by considering an optimal policy defined as

$$\pi_n^*(s, a) = \begin{cases} 1 & \text{if } a \in \arg\max_{a \in \mathbb{A}(s)} Q_n^*(s, a) \\ 0 & \text{otherwise} \end{cases} \quad (8)$$

Note that this policy is deterministic.

## 2.4 Reinforcement Learning

The mathematical foundations established in the previous sections serve as the basis for building algorithms to determine decision rules. In the field of RL, numerous algorithms aim to learn optimal policies. We have chosen to present two of these algorithms to illustrate a first distinction between online and offline application contexts. Furthermore, the second algorithm presented has been widely adopted to meet our application context. A discussion on the different RL algorithms suitable for our context will be the subject of Section 3.5.

### 2.4.1 Forward Q-learning

Q-learning, proposed in 1989 by Chris Watkins (Sutton & Barto, 2018; Watkins & Dayan, 1992), is one of the most famous and widely used algorithms in RL. It was historically developed in the so-called online context where the algorithm can dynamically interact with its application context. This is associated with the notion of 'agent', which is an entity capable of interacting with the environment while receiving rewards. The concept of interaction is related to the exploitation-exploration dilemma. The agent must, through trial and error, choose between exploiting acquired knowledge to maximise immediate rewards or exploring new actions to discover better long-term strategies (Sutton & Barto, 2018). An excellent illustration of this problem is the $\epsilon$-greedy strategy presented in the following definition:

**Definition 2.10.** *$\epsilon$-greedy Policy.*

$$\pi_\epsilon(s) = \begin{cases} \text{random action from } \mathbb{A}(s) & \text{with probability } \epsilon \\ \arg\max_{a \in \mathbb{A}(s)} Q(s, a) & \text{with probability } 1 - \epsilon \end{cases}$$

where $\epsilon \in [0, 1]$ is an hyperparemeter called the exploration rate.

Q-learning relies on the recursive Bellman equations 2.3. The idea is to estimate value functions based on the differences between current and previous estimates, and then to derive an optimal strategy from Equation 8 of Bellman optimality.

---
**Algorithm 1** Q-learning
---
**Initialisation** : $Q(s, a)$ arbitrarily, set learning rate $\alpha$, discount factor $\gamma$, and exploration rate $\epsilon$
**for** each history to build **do**
    Initialize state $s$
    **while** $s$ has not reached the terminal stage **do**
        Choose action $a$ using policy derived from $Q$ (e.g., $\epsilon$-greedy)
        Take action $a$, observe reward $r$ and new state $s'$
        Update $Q(s, a)$ using the Q-learning update rule:
            $Q(s, a) \leftarrow Q(s, a) + \alpha \left( r + \gamma \max_{a'} Q(s', a') - Q(s, a) \right)$
        $s \leftarrow s'$
    **end while**
**end for**
**Output**: The optimal decison rule is determined such as $\pi^*(s, a) = \arg\max_a Q(s, a)$
---

### 2.4.2 Backward Q-learning

When exploration of the environment is challenging, learning can be conducted using existing data, allowing decision rules to be derived from a non-interactive environment. This is referred to as offline or batch-RL. In this context, the algorithm does not interact with its environment; learning relies on estimating value functions from pre-existing databases. This offline Q-learning (Ernst *et al.*, 2005; Ormoneit & Sen, 2002) follows a backward approach illustrated in Figure 1.

The estimates of the Q-function are initialised at the terminal time and move backward in time step by step. This strategy allows for the consideration of a possible delay effect commonly observed in longitudinal data. To estimate the Q-functions, various regression algorithms can be used, such as linear regression, support vector machines, decision trees or by deep neural networks, among others.

---
**Algorithm 2** Backward Q-learning
---
**Input**: A set of training offline data consists of patients admissible histories $h_t$ and their associated indexed reward $r_t$, $t = 0, ..., \tau$ and a regression algorithm
**Initialisation** : Let $t = \tau + 1$ and $\hat{Q}_t$ be a function equal to zero everywhere on $\mathbb{S} \times \mathbb{A}$
**while** until $t = 0$ **do**
    $t \leftarrow t - 1$ (Backward)
    $Q_t$ is fitted with a regression algorithm though the following recursive equation : $Q_t(s_t, a_t) = r_t + \max_{a_{t+1}} \hat{Q}_{t+1}(s_{t+1}, a_{t+1})$
**end while**
**Output**: Given the sequential estimates of $\{\hat{Q}_0, ..., \hat{Q}_\tau\}$, the sequential optimal policies $\{\hat{\pi}_0, ..., \hat{\pi}_\tau\}$ can be determined
---

**Remark 2.10.** *In an offline context, direct exploration is not present because decisions are made based on data collected in the database. Although there is no longer an exploration-exploitation* dilemma as in the online context, it will be necessary to take into account a bias resulting from data where exploration-exploitation has already been performed.

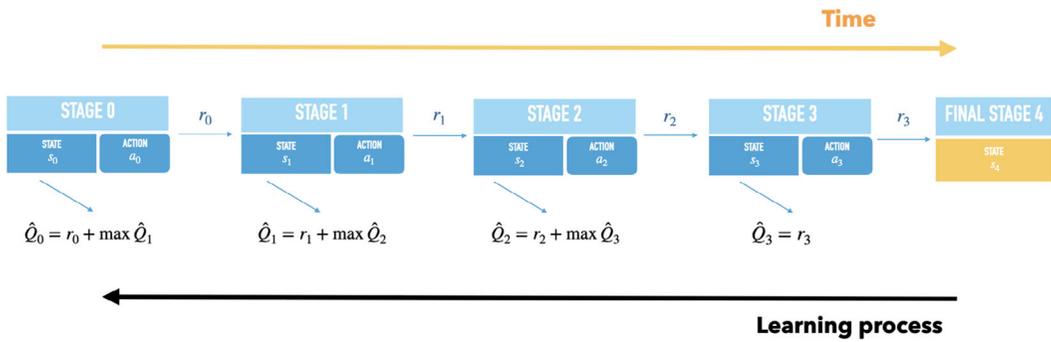

**FIGURE 1.** *Illustration of the Backward Q-learning algorithm for estimating Q-values on a history with four steps.*

## 3 Dynamic Treatment Regimes and Reinforcement Learning

### 3.1 Dynamic Treatment Regimes

Until the end of the 20th century, progress in medicine followed a 'one-size-fits-all' approach. The search for the effect of a treatment or intervention was framed within evidence-based medicine on a target population. With the advent of massive data, particularly genomics, the paradigm has evolved. The volume of individual data collected has exploded, suggesting the possibility of integrating individual factors in the search for the effect of an intervention. The desired effect of treatment is no longer an average effect but a conditional effect on patient characteristics.

In this context, where the effect of an intervention is conditional to the variable characteristics of the patient which vary over time, the relevance of a treatment for a given individual may also vary over time. A central objective of precision medicine is to develop adaptive, and potentially optimal, intervention rules, where the definition of optimality must be clearly defined (Kahkoska et al., 2022).

The search for adaptive (optimal) intervention rules is not a new question. A vast literature, primarily in the field of causal inference, exists and has real practical relevance. The foundational works in this context are attributed to Robins (1998), and the three extensions that allow for the effects of time-varying regimes in the presence of confounding variables: G-computation (Robins, 1986), the method of structural nested mean (Robins, 1994) models and G-estimation (Robins, 1992; 1989; 1998), as well as marginal structural models (Robins, 2000) and methods associated with inverse probability of treatment weighting (Chesnaye et al., 2022). Subsequently, a number of methods have been proposed, both in frequentist and Bayesian frameworks. All estimate the optimal DTR based on distributional assumptions of the data generation process via parametric models. We can consider them as direct resolution methods. These methods will not be further developed in this article; an up-to-date review including direct methods can be found in Deliu & Chakraborty (2022).

In the following section, we will detail the parallel that can be drawn between DTR and RL, which helps overcome a major barrier of direct methods, namely, the risk of misspecification of underlying assumptions (Zhao et al., 2015). To address this limitation, in Murphy (2003), followed immediately by Robins (2004), semi-parametric methods were considered, marking the first examples of RL-based approaches in the literature on DTR. The innovations of RL have breathed new life into the search for optimal DTRs, gradually expanding its applicability domain.

## 3.2 Decision Process and Dynamic Treatment Regimes

In Section 2, we notably introduced decision processes, policy and rewards which forms the theoretical foundation for algorithms searching for optimal policies, namely, RL. To describe the contribution of RL algorithms in the medical context, we will begin by examining how the framework introduced and DTRs are linked.

As discussed in Section 3.1, an adaptive intervention involves making a treatment decision based on the patient's characteristics and treatment history. An adaptive decision rule can thus be perceived as a policy in the theoretical sense presented in Section 2.2. To leverage the results of RL, it is essential to define the applied framework of the underlying DP for DTRs.

Building upon Definition 2.1 of a decision process, it is natural to consider, in a medical context:

- The state space $\mathbb{S}$ contains the selected covariates describing the patient's state.
- The action space $\mathbb{A}$ contains the selected treatments and their associated dosages.
- The subset $\{\mathbb{A}(s)|s \in \mathbb{S}\}$ states that the treatments feasible or accessible for a patient depend on a given state.

**Remark 3.1.** *It is worth noting that in our context, the variable $S_t$ is a vector containing a set of covariates observed at time t describing the patient's health state, which may influence the transition probabilities from one state to another.*

The observed histories $h_t$ are then the care pathways of different patients. They contain health data and treatments administered up to decision $t$.

One of the key elements of RL is the reward. In the medical context, rewards are defined to address the clinical objective. This is a very important point as optimisation relies on it. The notion of reward will be central in the discussion on the integration of medical expertise in Section 4. Indeed, for a given situation, different rewards can be associated depending on the expertise of the physicians, the specific objectives of the clinical trial, either proximally (directly after the decision) or distally (at the end of the follow-up).

## 3.3 Specificities of the Medical Context

DTRs find their primary application in medical contexts where multiple treatment lines are possible or in contexts with multiple possible decision points (see Figure 2). These adaptive strategies are particularly relevant in areas such as intensive care, chronic diseases, psychiatry or oncology.

The medical context is known for the great heterogeneity of its data (Kahkoska *et al.*, 2022; Sperger *et al.*, 2020), whether in terms of care pathways, treatment response, side effects, social factors or lifestyle. In this regard, data-driven methods offer promising perspectives by addressing the issue of model misspecification. Unlike traditional approaches that rely on pre-specified models, data-driven methods are based on real-world, observed data. This allows them to better reflect natural variations and provide more realistic insights. However, these methods depend heavily on the quality and representativeness of the data, and their results can be influenced by biases. Despite these challenges, data-driven approaches have the potential to improve precision medicine by enabling more accurate and personalised treatment recommendations. This could help reduce disparities and promote more equitable access to effective therapies.

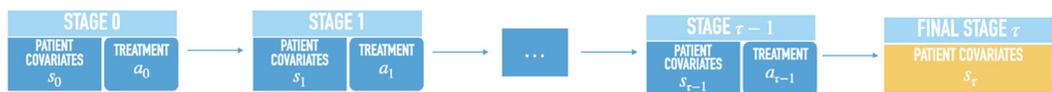

**FIGURE 2.** *Illustration of medical history: treatment line for a patient.*

The timing of decision-making moments is a central issue in the problem of adaptive interventions. Typically, these decision points are linked to patient visits to the practitioner. It is therefore natural to consider these moments as discrete and finite and to model them using a finite-horizon DP introduced in Section 2.1. Two issues arise: the time interval between two decision points and their frequency.

The issue of non-homogeneous time intervals between patients in the context of DTRs is typically addressed by considering the time between two visits as a covariate. Technically, this means defining the time based on the protocol and not worrying about the actual calendar time between visits. Even if the visits are not evenly spaced, by including this time information in a variable, we can treat the visits as if they are evenly spaced within the Markovian framework (Laber et al., 2014; Laber & Staicu, 2017; Schulte et al., 2014). Another important consideration is that the progression of time may not be uniform across patients, as some individuals experience much faster health deterioration or improvement than others. One possible approach to address this heterogeneity is to incorporate the patient's rate of evolution or health status directly into the state representation used in the decision process. By doing so, and by defining time steps relative to patient-specific health milestones rather than absolute calendar time, the Markov decision process framework can still be appropriately applied. There is, however, considerable room for research to develop RL frameworks that adapt dynamically to patients' evolving health states, an area that remains largely unexplored.

In some scenarios, such as patient follow-up in oncology or diabetes care, the number of visits is indefinite and varies based on individual patient needs. These patients are regularly monitored through mobile-Health (m-Health) initiatives, which operate in an online environment. Therefore, employing the Q-learning approach with backward induction, as explained in Section 2.4.2, becomes impractical. In Luckett et al. (2019), researchers identified optimal DTRs within an indefinite horizon framework using V-learning. This method aims to estimate an optimal policy from a predefined class of policies. Another approach, discussed in Ertefaie & Strawderman (2018), util an inferential procedure for estimating Q-functions.

**Remark 3.2.** *In the rapidly expanding field of m-Health research, online approaches are particularly suitable. Just-In-Time Adaptive Interventions (JITAIs) have already been the subject of research efforts (Istepanian* et al., *2007; Nahum-Shani* et al., *2018; Rehg* et al., *2017). A synthesis of JITAI research is provided in Deliu* et al. *(2024), along with a comparative study with DTRs. This study addresses the technical aspect of making decisions about adaptive treatments in an interactive online environment. We will not cover these aspects further in the work, as the framework of DTRs on observational data is discussed in Section 2.4.2, which is only feasible in the context of offline algorithms.*

## 3.4 Real Data Application

Data S1 provides an overview of the RL research conducted in the context of DTRs. It is important to note that decision points are typically few in observational data application context;

many studies consider two or three decision points. This choice is primarily driven by computational challenges: the more decision points there are, the more complex it becomes to integrate the patient's history into the models. An alternative approach is to impose a Markov assumption on the DP. However, in healthcare applications, this assumption is often unrealistic. The entire patient history can rarely be ignored or encapsulated in the current state.

As with any analysis on healthcare data, it is natural to question the biases inherent in the methods and the issue of causality (Hernan & Robins, 2023; Neuberg, 2003). Since machine learning techniques are not causal inference methods, their use requires data that satisfy causal assumptions. However, in the absence of a well-defined causal structure, such methods may capture statistical associations that do not reflect underlying causal mechanisms. The issue typically arises in terms of 'potential outcomes', and it is common to consider causal inference assumptions such as the 'stable unit treatment value' assumption and the 'no unmeasured confounders' assumption, as explained in chapter 2 of Chakraborty & Murphy (2014). The question of causality in the field of RL is also addressed more directly in the framework of 'causal RL'[1] (Chakraborty & Murphy, 2014; Zhang, 2020). The search for adaptive intervention rules relies on data with a specific longitudinal structure. Innovations in algorithms for finding optimal DTRs often begin with adjustments to existing observational databases.

The Medical Information Mart for Intensive Care (MIMIC) (Johnson *et al.*, 2016) is a publicly accessible observational database containing information on 53,423 distinct admissions for patients in intensive care units between 2001 and 2012. It includes data on vital signs, medications, laboratory tests, measurements, caregiver notes, procedure and diagnostic codes, imaging reports, length of hospital stay, survival data and so on. Due to the wealth of available information and its longitudinal nature, MIMIC has been widely used by the RL community as a support for methods comparison (see (Roggeveen *et al.*, 2021), Table 1 and Data S1). It is also utilised as a training dataset for the development of data augmentation methods (Tseng *et al.*, 2017) and the generation of interactive environment models (Peng *et al.*, 2018; Raghu *et al.*, 2017).

Similarly to how randomised trials play a distinct role in clinical research and may be considered the gold standard for causal relationship investigation, the sequential multiple assignment randomised trial (SMART) design (Cheung *et al.*, 2015; Kosorok & Moodie, 2015) can be regarded as the gold standard for clinical trial design in the context of adaptive interventions. SMART designs involve an initial randomisation of patients to various treatment options, followed by re-randomisations at each subsequent stage of some or all of the patients to another available treatment at that stage. With such a design, the stable unit treatment value assumption is 'by design' fulfilled. However, SMART designs are challenging to implement, costly, and as a result, there is limited access to data from SMARTs. However, notable trials include

- CATIE (Clinical Antipsychotic Trials of Intervention Effectiveness) is a SMART study involving 1,460 schizophrenia patients over 18 months aimed at evaluating the clinical effectiveness of specific sequences of antipsychotic medications (Shortreed *et al.*, 2011).
- ADHD (Attention Deficit Hyperactivity Disorder) is a SMART study involving 150 simulated participants, aimed at evaluating an adaptive intervention for children with this disorder. This study integrates behaviour modification treatment along with medication treatment (Chakraborty & Murphy, 2014; Laber *et al.*, 2014).
- STAR*D (Sequenced Treatment Alternatives to Relieve Depression) is a SMART study involving 4,041 patients with major depressive disorders. This study evaluated the effectiveness of different treatment regimens (Chakraborty & Murphy, 2014; Laber *et al.*, 2014).

Table 1. *Applications of RL algorithms on MIMIC database: highlighting various medical objectives with rewards design extract from (Roggeveen* et al., *2021).*

| Reference | Model | State space | Action space | Rewards |
|---|---|---|---|---|
| Komorowski et al. (2016) | SARSA | Discretised state space | 25 unique actions based on a 5 by 5 binning procedure of maximum vasopressor dose and sum of intravenous fluids per 4h time interval | Terminal reward at the end of each trajectory based on 90-day mortality |
| Raghu et al. (2017) | Duelling DDQN | Ordinary and Sparse Auto-Encoders were used for latent state space representation | As paper Komorowski et al. (2016) | Terminal reward at the end of each trajectory based on in-hospital mortality |
| Raghu et al. (2017) | Duelling DDQN | Continuous state space based on 4h aggregated features based on physiological parameters | As paper Komorowski et al. (2016) | Intermediate reward based on changes in critical care scores and lactate combined with a terminal reward for survival based on ICU mortality |
| Peng et al. (2018) | Duelling DDQN | Patient states are encoded recurrently using an LSTM autoencoder representing the cumulative history for each patient | As paper Komorowski et al. (2016) | The change in the negative mortality logodds of mortality between the current observations and the next observations. |
| Li et al. (2019) | Actor-Critic | POMDP | As paper Komorowski et al. (2016) | As paper Komorowski et al. (2016) |
| Yu et al. (2019) | Duelling DDQN | As paper [3] | As paper Komorowski et al. (2016) | Developed several reward functions based on 7 potential features most important during the treatment process |

## 3.5 Properties of RL Applied to DTR

There is a wide range of RL algorithms offering various methodological approaches tailored to specific contexts, as illustrated in Table S1. Figure 3 below provides a non-exhaustive overview of the most common RL algorithms. It presents many dichotomies, which will be explained in the following paragraph and contextualised in DTRs applications.

### 3.5.1 Model-based vs. Model-free

The first dichotomy in Figure 3 is based on the distinction between a model-based approach and a model-free approach. This distinction is related to the concept of transition probability defined by Equation 1. A procedure is considered 'model-based' when it relies on knowledge of all transition probabilities from a model, which means having access to all dynamics of the system. A model-free method is able to bypass this model and is based on partial information of the associations between states and actions to determine the optimal strategy. In a model-based approach, all possible paths from an initial state $s_0$ are explored, and an optimal policy is one that maximises the objective.

However, in a medical context, exploring all possibilities from the same starting point is infeasible, mainly for clinical and ethical reasons. The environment is thus inherently partially observed. This reality inherently places us in a model-free framework. It worth noting the existence of an application on simulated patient data based on MIMIC (see Section 1) in the model-based framework in Raghu et al. (2018).

### 3.5.2 Policy-based vs. Value-based vs. Actor-critic

The second distinction involves two different approaches to determine the best strategy: policy-based methods and value-based methods. The former aim to directly find an optimal policy by formalising the RL problem through a family of policies, introduced in chapter 13 of Sutton & Barto (2018). The latter seek an optimal policy through value functions, introduced in Section 2.3.2, and serve as the basis for algorithmic methods such as dynamic programming, Monte Carlo, and temporal-difference, also presented in the same book. These two approaches can be combined, thus forming actor-critic methods (Grondman *et al.*, 2012; Sutton & Barto, 2018).

*Policy-based*   Policy-based methods are direct approaches to finding an optimal policy that rely on a parametric form of the strategy $\pi_\theta$ for $\theta \in \Theta$. Optimisation can be typically achieved through gradient descent:

$$\theta_{n+1} = \theta_n + \nabla \mathbb{E}_{\pi_\theta}[G_n | \theta] \qquad (9)$$

where $G_n$ is the cumulative long-term reward introduced in Remark 2.8.

This method has been applied to simulated HIV data (Yu *et al.*, 2019) as well as in the intensive care domain (Raghu *et al.*, 2018). Note that the first application highlighted the challenges of converging to an optimal decision rule due to the simplification of simulation models. The main obstacle to using this method is the difficulty of convergence, which requires a large volume of data.

*Value-based*   Value-based methods evaluate an optimal policy indirectly based on value functions $V^\pi$ or $Q^\pi$ introduced in Section 2.3.2. The general idea is to quantitatively evaluate states or action-state pairs using one of the value functions (Q-function or V-function). An optimal policy is then obtained by identifying actions that maximise these values. The success of these methods relies on the ability to model these value functions, as outlined in Section 2.4.2, through algorithms such as Backward Q-learning, making it a highly flexible approach.

The initial work was conducted by Murphy (2005b), who introduced an offline Q-learning, also known as batch learning, in a context of non-Markovian planning with a limited and restricted number of steps ($n \leq 4$). This approach proves ideal for its application to DTRs and can serve as a starting point for many other applications. Research activity in this field quickly became significant, considering various parametric, semi-parametric, and non-parametric strategies to model the value function (Chakraborty & Murphy, 2014; Laber *et al.*, 2014; Murphy, 2005a; Tsiatis *et al.*, 2019).

Value-based methods are better suited for application to DTRs. They enable the discovery of optimal decision rules in a non-Markovian framework with a small number of steps and data, unlike policy-based methods. This makes them easily applicable to observational data. Moreover, they can offer a clearer interpretation, especially when Q-function estimation relies on a linear regression model (Laber *et al.*, 2014), thus providing interpretable decision rules. As shown in Data S1, this is the most widely used method in practice, particularly Q-learning approaches and its derivatives in the context of DTRs.

*Actor-critic*   A third approach to address the question of finding an optimal strategy is known as the 'Actor-Critic' method. It takes a hybrid approach by combining an actor based on policy-based methods with a critic based on model-based methods, thus integrating the advantages of both previous methods. The actor refines the parameterised policy under the guidance of the critic. The latter uses value functions, also parameterised $V^{\pi_\theta}$ or $Q^{\pi_\theta}$, to guide learning. This third way of constructing decision rules was developed to correct biases in value-based methods and to counterbalance the high variability of the gradient part of policy-based methods in Equation 9 (Grondman *et al.*, 2012).

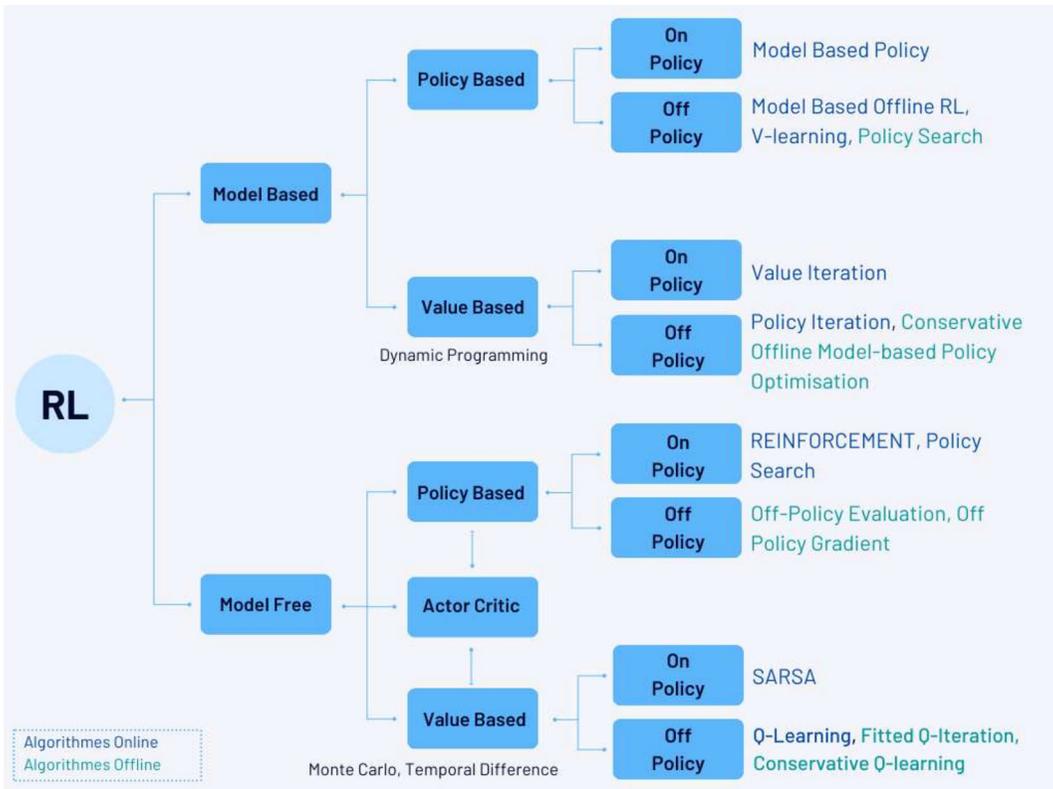

**FIGURE 3.** *Classification of the most common RL algorithms.*

Actor-critic methods have been applied to the MIMIC dataset. This compromise between policy-based and value-based methods converges towards a decision rule reducing patient mortality in Wang *et al.* (2018) or providing a decision rule in line with physician's usual opinions in Li *et al.* (2020) and Li *et al.* (2018). This approach relies on gradient descent, similar to policy-based methods, thus necessitating databases containing a large number of individuals, often simulated data.

### 3.5.3 On-policy vs. Off-policy

This dichotomy is closely related and sometimes confused with the concepts of offline and online algorithms presented in the following Section 3.5.4.

DTRs on observational data inherently operate in an offline context, aiming to determine an optimal policy using previously collected data. This means that, rather than adapting in real time, the analysis and optimisation are done retrospectively. For example, all SMART designs rely on data gathered from established clinical protocols or previously collected observational studies. These protocols dictate the timing and nature of patient visits, ensuring a structured collection of data. By analysing this data, researchers can develop and refine treatment strategies (Uehara *et al.*, 2022; Chakraborty & Murphy, 2014; Kosorok & Laber, 2019). Therefore, applying RL in the DTR context and clinical decision support is fundamentally off-policy, meaning that the strategy used to generate the data ('behaviour policy') is not necessarily optimal. The optimal strategy ('target policy') is deduced subsequently.

On-policy algorithms require an interactive online context where the strategy generating the data is optimised. The concepts of behaviour and target policies are merged. The online framework can benefit from both on-policy algorithms, as is the case in the medical domain with Just-in-Time Adaptive Interventions (JITAIs) discussed in Deliu *et al.* (2024), and off-policy algorithms (see Figure 3). Some online algorithms, both off-policy and on-policy, have been explored within the context of DTRs, but exclusively in simulated data settings, as indicated in Data S1.

### 3.5.4 Online vs. offline

The distinction between online and offline settings is fundamentally determined by the nature of data collection and whether interaction with the environment is feasible at the time of decision-making. When interaction is possible, the setting is considered online; otherwise, it is offline.

In the medical context, allowing an algorithm to interact directly with patient care raises significant ethical concerns, particularly regarding the trial-and-error nature of learning explored by online algorithms as seen in Section 2.4.1. Indeed, in many clinical scenarios, it is unacceptable to assign treatments randomly or in a potentially suboptimal manner, which severely limits the scope for active exploration.

In precision medicine, DTR offer a formal framework for personalising care over time by adapting treatments to patient responses. In offline settings, where learning is done from existing datasets, the gold standard for data collection is the SMART. These trials are specifically designed to evaluate multi-stage interventions by incorporating planned re-randomisations based on intermediate outcomes. A detailed example of such a design is described in (Chakraborty & Murphy 2014, chapter 2). In this SMART for addiction management, individuals are initially randomised to receive either cognitive behavioural therapy (CBT) or naltrexone (NTX), a medication used to treat alcohol dependence. After two months, patients are classified as responders or non-responders based on whether they have had more than two heavy drinking days. Non-responders to NTX may be re-randomised to either switch to CBT or receive an augmented treatment combining NTX, CBT, and enhanced motivational support. Similarly, non-responders to CBT may switch to NTX or receive the same augmentation. Responders, meanwhile, may be re-randomised to receive either telephone monitoring (TM) or telephone monitoring plus counselling (TMC) for an additional six months. This design allows for evaluation of various treatment sequences, with the aim of maximising the number of non-heavy drinking days over a 12-month period. The design of such studies requires substantial involvement of medical experts, as both the timing of treatment adaptation and the set of available treatment options must reflect clinical practice and the current state of clinical equipoise, ensuring that findings are not only scientifically valid but also ethically and clinically relevant.

However, still within the scope of precision medicine, an online setting is also possible—particularly in the context of mHealth, as mentioned earlier, and implemented through JITAIs. For example, a smoking cessation mobile application (Yang *et al.*, 2023) can start with a 'warm start', using a set of randomised actions that are safe and generally effective for patients with given baseline characteristics—such as motivational messages, cognitive-behavioural prompts or mindfulness exercises. These initial choices should be guided by clinical expertise, and patients should be encouraged to follow their physician's recommendations. As more data become available over time (e.g., self-reported cravings, geolocation indicating proximity to usual smoking spots or wearable sensor data), the app can gradually learn which actions are most beneficial for each individual and adapt accordingly—for example, pushing a mindfulness video before a high-risk moment like a commute home. It is nevertheless important to maintain some degree of exploration in this process, in order to continue improving and to respond to relevant changes in the patient's condition and environment.

## 4 Integrating Medical Knowledge Into Reinforcement Learning Models

The previous section has highlighted the variety of algorithms available for seeking optimal decision rules. Regardless of the method used, the construction of decision rules remains algorithmic and data-driven. Therefore, the legitimate question arises regarding the explainability of the obtained decision rule, both for the patient and the practitioner. A prerequisite for the clinical application of these decision rules will be to address these concerns. To do so, we will explore how medical expertise can intervene in the construction of these decision rules.

This section has two main goals: firstly, to outline how medical knowledge intersects with RL algorithms in the search for treatment decision rules, and secondly, to propose adjustments to these algorithms to better suit their application to DTR. These twin aims are aimed at enhancing the safety, interpretability, and relevance of tools for medical decision-making.

### 4.1 Medical Knowledge and Model Preparation

Like any machine learning method, the search for the optimal DTR depends on the data from which the method was trained. Data preparation is therefore an essential step. Medical knowledge is certainly involved in this process. Indeed, in this causal context, the choice of variables to collect and the selection of confounding factors are crucial. These decisions are primarily guided by medical expertise, drawn from the experience of practitioners and medical literature, as detailed in Section 3.3. The construction of the training dataset is thus the very first intervention of medical knowledge in RL models. It is primarily a methodological consideration that may bias the constructed optimal decision rule (Remark 2.10).

The second step in the preparation phase of applying RL in the context of searching for optimal DTRs involves selecting an algorithm from the various possibilities presented in Figure 3. This choice is primarily based on how the data were collected, the chronology of events, juxtaposed with the different characteristics of RL algorithms discussed in Section 3.5. The choice of method thus depends mainly on the application context and available data and, therefore, on underlying medical knowledge. Again, this is primarily a methodological issue, where the medical specialist collaborates with the machine learning specialist to make this choice or develop a new ad hoc method. This discussion could follow the decision tree outlined in the figure titled 'Overview of the guideline for the application of RL to healthcare' in Coronato et al. (2020).

### 4.2 Medical Knowledge and Rewards

One crucial aspect of learning optimal strategies is the formulation of rewards. This is a key component and one of the primary mechanisms for integrating medical knowledge into RL methods. In practical terms, commonly, the choice of reward is directly based on medical expertise. It is primarily a methodological issue closely linked to the definition of the study's objective. The selection of the reward is similar to choosing the primary outcome in the design of a clinical trial, with the same imperatives of precision and representativeness of the variable.

Rewards mainly consist in scores or quantitative variables, such as changes in body mass index in weight loss studies (Linn *et al.*, 2015), or survival functions in critical care settings (Roggeveen *et al.*, 2021). Additionally, more complex rewards can be found, such as compromises or combinations of variables, as seen in oncology contexts (Zhao *et al.*, 2009), where the reward is evaluated considering tumour size, treatment toxicity, patient well-being, and survival rates. In Table 1, an illustration of various reward functions is provided, each aiming to achieve a specific medical objective.

It is evident that selecting an ad hoc reward for the problem under study can entail choices that are either too arbitrary or too context-specific, potentially leading to overly restrictive learning objectives. An alternative approach is to replace this choice of reward with reward shaping. Several approaches have been developed in this direction.

One way to generalise and automatically construct rewards is through inverse RL. This method uses patient trajectories generated with expert medical decision-making to extract an estimate of the underlying reward function for these choices. Thus, it also seeks to highlight the characteristics that should be considered for its formulation. The latent medical knowledge will then be encapsulated in the estimation of the reward function. This approach has been used in the context of alcohol addiction management (Shah *et al.*, 2022) for the search for a personalised decision-making rule. The application of inverse RL to the framework of DTRs is also explored in the article (Luckett *et al.*, 2021), where the objective of this study is to construct a reward function as a linear combination of covariates. In Perera *et al.* (2025), a method for preparing data for reward function inference is proposed. First, an expert or an oracle categorises the decisions of trajectories into three categories: optimal, sub-optimal, and non-optimal. The reward function is then learned through inverse RL, based on the idea that trajectories generated by policies classified as optimal should receive the highest reward. Once these rewards are established, the learning of an optimal policy is achieved through Deep Q-learning. This article presents applications in the context of sepsis (MIMIC) and diabetes. Inverse RL allows for the determination of rewards from data, thereby accelerating the learning of a decision rule compared with manually constructed rewards. It is important to note that these methods assume that the physicians who generated the training data made decisions aimed at maximising the interests of each patient. Thus, the constructed rewards are sensitive not only to the quality of the data but also to medical decisions. Moreover, inverse RL can help address a key limitation of observational data, namely, that under-represented subgroups may lack sufficient data to reliably learn treatment effects or optimal policies. By learning from clinicians how they prioritise multiple outcomes, such as measures of efficacy, measures of side effects, and measures of cost, inverse RL constructs an expert-informed composite utility function. This utility function can then be incorporated into RL algorithms to more effectively optimise patient health outcomes.

Another way to generalise the construction of rewards based on expert knowledge is preference learning. A subfield of research in machine learning, it relies on the idea that the expert provides preferences between two elements, which induces a ranking among these elements. Combined with RL, preference learning uses this induced ranking to guide the policy learning. In a model-based and online framework (Fürnkranz *et al.*, 2012), preference learning replaces rewards to induce a preferred action based on preferences between trajectories, states or policies. The principle is to use a simulation model to generate all possible trajectories from all possible actions, then select the preferred ones using a preference model. In an online, model-free, and off-policy framework, learning an optimal strategy is done in three steps (Akrour *et al.*, 2012). First, an exploratory phase where trajectories are generated by a behaviour policy. Second, an expert provides preferences, which induces a ranking. Third, the model learns an optimal strategy by solving a constrained optimisation problem where the preferences are modelled within the constraints. In an offline framework, preference learning separately learns the rewards and the optimal strategy. The comparisons are then used in a probabilistic model, such as the Bradley-Terry algorithm, to construct rewards by maximum likelihood estimation or neural networks (Shin *et al.*, 2022). These rewards are then integrated into RL algorithms.

Preference learning methods, described as model-based/on-policy by Fürnkranz *et al.* (2012) and model-free/off-policy by Akrour *et al.* (2012), use preferences on trajectories on simulated data similar to the generic cancer scenario described by Zhao *et al.* (2009). In Fürnkranz *et al.* (2012), patient trajectories are compared using a partial order relation that considers survival, maximum toxicity over time, and final tumour size. Meanwhile, (Akrour *et al.*, 2012) formulate expert preferences by prioritising trajectories with superior final outcomes, which include minimal tumour size and reduced toxicity levels. Preference Learning enables the construction of rewards based on expert preferences on trajectories, allowing learning to rely on explainable choices. However, the applications described in the articles (Fürnkranz *et al.*, 2012; Akrour *et al.*, 2012) are based on simulated cancer data and simulated preferences and have been developed in an online framework, which is not suitable for direct clinical application. An offline, off-policy solution is proposed in Shin *et al.* (2022), but it has been developed in the context of robotic or video game applications.

Other methods for constructing rewards exist, such as human-centred RL, which utilises rewards directly provided by an expert. The agent interprets expert feedback as numerical rewards. These approaches are detailed in Li *et al.* (2019), but they are generally applied in an online and on-policy context, which involves direct interaction of the agent with patients, thus raising ethical concerns and requiring a specific application framework beyond the scope of this article.

## 4.3 Medical Knowledge and Value Functions

The evaluation or estimation of value functions $V_n^\pi$ and $Q_n^\pi$ is also a key concept in RL. In the medical context, due to the complexity of environments and the volume of available data, these assessments often suffer from a lack of precision. Integrating medical expertise can be considered to improve results.

This is particularly true when medical expertise translates into knowledge of treatment response mechanisms. Indeed, these observations can then be integrated into RL methods to guide the learning of the optimal strategy. From a technical standpoint, it is conceivable to penalise the value function: decrease the value function when mechanisms identified by an expert indicate that the treatment is inappropriate and increase the value function when the treatment is considered relevant. Actions associated with a lower value function are less likely to be selected than those associated with a higher value function. This approach thus highlights actions considered more relevant by the expert and guides learning in the right direction. This approach was implemented for patients with renal failure in Gaweda *et al.* (2005). Medical experts identified that patients who do not respond to standard treatment require higher doses. The authors constructed a DTR by incorporating this clinical fact into a Q-learning algorithm. When a patient does not respond to a treatment dose, the Q-values of lower doses are penalised, thus favouring higher doses. This approach offers the advantage of reducing the need for exploration and hence the learning time. However, it was developed in an online framework using simulated data, limiting its applicability to observational data.

The integration of medical expertise can also occur through relay collaboration. The principle involves considering two concurrent value functions: $Q$, the usual value function, and $Q^{clin}$, the value function under the practitioner's strategy in a given situation. The latter comes into play only when the patient is in a critical state, as evidenced by their vital signs. Subsequently, this decision and the patient's response to treatment will be used to enrich the learning model through an enhanced value function, denoted as $Q^+$. Thus, the strategy for updating the value functions involves recommending treatments suggested by the RL model while seeking the expertise of physicians when the patient's condition is deemed critical. $Q^+$ can therefore be formalised as

$$Q^+(s_t, a_t^+) = \begin{cases} Q^{clin}(s_t, a_t^{clin}) & \text{If the patient's covariates indicate a critical state} \\ Q(s_t, a_t) & \text{Otherwise} \end{cases}$$

where $a^{clin}$ is the treatment chosen by the clinician.

This approach has been deployed in the context of intensive care treatment in Wu *et al.* (2023) when the patient exhibits severe symptoms. In such situations, RL algorithms may propose aggressive treatment strategies to maximise reward, which can entail significant risk for the patient. In this study, a model based on value functions $Q$ incorporates human expertise on the treatment of sepsis. Applied to the MIMIC database, this model is evaluated using a score reflecting the patient's critical state. Expert intervention is triggered when the score is considered low. The application of this method demonstrates a higher survival rate compared with some similar methods without human expertise and also improves the estimation of the value function.

The principle of collaboration between the agent and the expert is also addressed in the article (Sonabend *et al.*, 2020) using the MIMIC database. It still impacts the $Q$-functions, but now through a statistical test. The idea is to introduce exploration into an offline model by comparing risks between two strategies. One simulates standard medical decisions, while the other strategy suggests an alternative treatment. From a comparison test on state values associated with a policy, one of these strategies is adopted. The question is: when could a new treatment be better than conventional therapies? The solution seeks to balance choices of standard treatments with new options while assessing risks to discover promising alternatives that physicians have not considered.

This connection between RL and medical expertise enables both the supervision of treatments in complex cases and the exploration of alternative approaches while assessing associated risks. Although off-policy RL methods can be prone to data biases, they have the potential to enhance medical practice by integrating data-driven insights with physicians' perspectives. Integrating medical knowledge involves either analysing health data to observe medical mechanisms or incorporating direct input from physicians. In either case, this integration must strike a balance between data, expert opinion, and statistical models to determine the most suitable treatment. Errors may arise from both the expert's judgement and the data collection process, highlighting the need for a nuanced approach that carefully considers and balances these potential sources of inaccuracy. Addressing these errors requires a comprehensive evaluation of data quality and expert validity, ensuring that the decision-making process is robust and well-informed. By doing so, we can more effectively minimise inaccuracies and improve treatment outcomes.

### *4.4 Medical Knowledge and Objective Function*

Value-based approaches can benefit from the integration of medical expertise in determining optimal strategies. Similarly, methods for incorporating medical expertise have been proposed for policy-based approaches, which directly modify on the objective function.

Supervised RL merges two subfields of machine learning: supervised learning and RL. The fundamental principle of this method is to maximise a long-term objective, with the supervision of an expert, in order to maintain consistency with clinical treatment standards. Its ultimate goal is to predict an optimal treatment policy, minimising deviations from medical expert recommendations. In this framework, the expert plays a crucial role as a reference for training the RL algorithm, using a database containing all medical decisions made within a cohort. This control affects the objective function in two ways. The latter is simplified into two parts: the first, derived from an actor-critic algorithm, aims to perfectly mimic the experts through its 'critic' part (Section 3.5.2.3). The second part of supervised learning minimises the difference between predicted treatments and those traditionally administered. This method, described notably in Yu *et al.* (2020), is applied in the intensive care domain using the MIMIC database and focusing on ventilation and sedation dosing. The primary objective is to provide optimal care that respects both short-term and long-term goals for patients, while adhering to best clinical practices. In this context, research shows that the supervised RL approach outperforms the classical Actor-Critic approach in terms of convergence speed and alignment with usual medical decisions. In the study by Wang *et al.* (2018), the supervised RL approach was applied to the MIMIC dataset. The treatment recommendations obtained would lead to a decrease in patient mortality rates. Supervised RL, in its fundamental construction, aims to perfectly mimic the usual treatment practices, making it an excellent means of emulating practitioners. However, it prevents for the proposal of alternative or less explored treatments compared with usual care methods.

*4.5 Medical Knowledge and Policy*

It is important to note that medical decision rules constructed within the framework of RL recommend only a single action for a given state. The multiple policies approach involves proposing different equivalents or closely related strategies for a given patient state. Consequently, the specialist, relying on their expertise and the constraints of their environment, chooses the treatment from the selection of actions offered. This approach introduces the notion of quasi-equivalent actions that may take into account considerations such as side effects, less invasive treatments, and local availability. Essentially, the general idea is to train a set of policies evaluated by value functions, which learn a correspondence between each state and a collection of closely comparable actions. Subsequently, the approach involves restricting the choice of actions by evaluating the extent to which the deviation from optimal value is acceptable. This is the concept of worst-case value, referring to the expected gain in the worst possible scenario within the set of allowed actions. The level of deviation from optimality allowed will be controlled by a hyperparameter.

The concept of multiple policies was introduced in Milani Fard & Pineau (2011) and applied in a simulated setting of sequential clinical trials for patients suffering from depression. It was developed within a model-based, on-policy, online framework with a finite horizon, not conducive to observational data or real clinical applications. In the article (Tang *et al.*, 2020), the method evolved into a model-free and off-policy framework, still online using the temporal difference learning algorithm, and was applied in the simulated context of critical care based on MIMIC. Like the previous method, its development in an online environment does not align with our application context, but it establishes the foundation for a model-free approach, thus representing progress towards a model suitable for DTR.

In conclusion, the concept of multiple policies has also been employed in a multi-objective context, not based on expert opinion but on patient preferences, as detailed in Lizotte & Laber (2016). By combining the notion of equivalent strategies with a multi-objective framework and Pareto dominance, and considering the preferences of patients, less restrictive solutions can be obtained. This approach, applied in the CATIE study specifically tailored to the DTR context, offers decision-makers increased choice by a larger class of optimal policies. These could provide the basis for an application that integrates experts' preferences and medical knowledge, thus addressing the issue outlined in this article.

## 5 Conclusions

This paper introduces and aims to facilitate the understanding of RL methods for precision medicine, especially its application to optimal DTR research. This topic is of major practical interest since it aims to determine an optimal decision rule for personalised treatments, with a large range of applications in areas such as intensive care, chronic diseases, psychiatry, and oncology. However, applying RL to medical research requires specific considerations and adaptations.

The main specificity arises from the data, typically derived from observational studies, which limits RL methods to offline applications. While an online setting is feasible, such as in m-health scenarios, for many cases, it is unethical to base treatment decisions solely on an algorithm. Therefore, since the data has already been collected beforehand, it is important to note that the well-known exploration-exploitation dilemma of online RL translates into an exploration-exploitation bias in offline RL settings. Section 3.5 details the properties of RL algorithms and helps identify the most desirable characteristics for an algorithm applied to DTR. First, due to clinical and ethical constraints, exploring all possibilities from the same starting point is impractical, necessitating the use of model-free algorithms. Secondly, value-based methods enable the discovery of optimal decision rules in a non-Markovian setting with limited steps and data, distinguishing them from policy-based approaches. Thirdly, off-policy algorithms are suited for offline contexts where data is already collected following a specific strategy, allowing for the determination of an optimal policy in a second phase. When these three characteristics converge, the result is an algorithm well suited for practical applications with observational DTR data. Consequently, Backward Q-learning, also known as Fitted Q-Iteration, emerges as the most widely adopted and utilised algorithm in the realm of applying RL to DTR (Clifton & Laber, 2020).

Intimately linked to all work on observational data, the question of causality arises in the optimal DTR research context. A few research works directly focus on this challenge (Chakraborty & Murphy, 2014; Zhang, 2020), but most of the time causality is based on assumptions that are difficult to verify which make the results questionable. This limitation may be overcome by the experimental design relying on SMART designs but such designs are difficult and expensive to set up (Cheung *et al.*, 2015; Kosorok & Moodie, 2015).

The classical formulation of RL relies on decision processes theory under the Markov assumption. However, this assumption is often too stringent in practical applications. Indeed, there is no guarantee that the current state under study contains all the necessary information to construct a precise decision. However most of the mathematical properties remain true without this Markov assumption by considering the entirety of the patient's history. In practice, that necessitates huge computational capacities and restricts to the applications the determination of adaptive strategies where the number of DTR steps is small (less than 4).

In addition to the previous issues, another problem emerges in the search for an optimal treatment strategy: the acceptability of the optimal DTR to both patient and practitioner. This raises concerns about how understandable the decision rules are for both patients and physicians, which is crucial for their clinical use. Integrating medical expertise into machine learning methods for personalised treatments is essential to improve safety, interpretability, and effectiveness in real-world scenarios.

One way to overcome this issue is to consider algorithms involving, one way or another, medical expertise or knowledge. We have seen that various approaches and studies demonstrate how medical expertise can be integrated into RL methods for sequential treatment decisions. This integration can be done at various stages of algorithm implementation.

First, the medical knowledge is often integrated before the study, at the design of the experiment. Indeed, physicians contribute to selecting the variables used for learning the decision rule. Similarly, algorithm selection involves collaboration between medical and machine learning expert, based on the application framework and available data.

Second, the medical knowledge can be integrated by acting on the rewards. Rewards is one of the main elements of a RL algorithm. Since they influence and guide the determination of the decision rule. Their design is thus crucial. Traditionally, a variable representative of the study's objective is chosen. Methods such as inverse RL and preference learning attempt to generalise their construction through expert input. Preference learning (Fürnkranz et al., 2012; Akrour et al., 2012) and human-centred RL (Li et al., 2019) directly incorporate expert knowledge into reward construction. However, this method suffers from being developed only in an online setup, which is not applicable to DTRs and observational clinic application. Nonetheless, early research in this area can serve as a foundation for further exploration. On the other hand, inverse RL is promising since it is developed within the offline context and is well suited for real clinical applications (Shah et al., 2022; Luckett et al., 2021). However, some applications require substantial training datasets to converge to a solution and rely on the Markov assumption, as shown in Perera et al. (2025).

Thirdly, the learning of decision rules can be achieved through value functions, allowing for the integration of medical expertise at this level. One approach is to incorporate observed medical mechanisms; specifically, the idea is to penalise the Q-values associated with non-decisive treatments (Gaweda et al., 2005). However, this method was initially developed in an online context and requires reassessment for offline settings. A second idea is to establish a relay between human decisions and decisions proposed by the algorithm. In one scenario, the physician would take over when the patient is in critical conditions (Wu et al., 2023). In another scenario, the algorithm would suggest alternative treatments to those traditionally proposed, along with associated risks (Sonabend et al., 2020). These hybrid methods seem promising for real clinical applications, but concrete evidence of their implementation is currently lacking. In the policy-based methodological framework, the integration of expertise can occur through a method called supervised RL (Yu et al., 2020; Wang et al., 2018). Its aim is to faithfully replicate common medical practices, offering precise emulation of physicians' decisions. However, it does not allow for the discovery of alternative or underexplored treatments compared with conventional care methods.

Lastly, the learning of decision rules can be approached methodologically through policy and it is worth noting that classical RL methods typically recommend only one policy, typically one treatment and one dose for each decision time. To enrich the context, multiple policies methods have been developed with the aim of offering an expert multiple equivalent treatment to choose from. The work of Lizotte & Laber (2016) is particularly suitable for application to observational data-based DTRs, but it was developed within a framework of patient preferences and could be reassessed within an expert preference framework.

RL and artificial intelligence (AI) in the service of humanity raise complex ethical questions, particularly in medicine. AI alignment (Gabriel, 2020) aims to ensure that AI systems act in accordance with human intentions and values, avoiding harmful or dangerous behaviours. However, learning from non-representative data can introduce biases, compromising the fairness of care. In DTRs, the coordination between AI recommendations and medical oversight presents challenges in accountability. Integrating medical expertise into algorithms could one approach to improve their ethical alignment, but this requires interdisciplinary collaboration to ensure fair and responsible systems.

The integration of medical knowledge is a promising research field, exploring various innovative perspectives and methods. However, further research is needed to adapt them to the specific constraints and realities of precision medicine. These advancements have the potential to lead to practical clinical applications and significantly enhance daily hospital operations. This aligns with the broader challenge of applying mathematical solutions effectively in clinical practice. Particularly, the development of Health System Science enables the use of interdisciplinary skills to study the complexity of healthcare systems (Apostolopoulos *et al.*, 2020; Kahkoska *et al.*, 2022). Practically, the aim is to ease the transition of laboratory discoveries into clinical practice (Gilliland *et al.*, 2019), achieved by forming interdisciplinary teams within healthcare systems. Combining progress in both research areas could establish a framework for applying RL alongside medical expertise, simplifying the treatment decision process for all parties involved. We hope this study will foster collaboration between machine learning researchers and healthcare professionals, by showing a framework that helps practically applying RL in DTR contexts.

## Conflict of Interest

The authors declare no potential conflict of interests.

## Endnote

[1]For details of 'causal RL' initiative, see https://crl.causalai.net/.

## References

Akrour, R., Schoenauer, M. & Sebag, M. (2012). April: Active preference learning-based reinforcement learning. In *Machine learning and knowledge discovery in databases: European conference, ECML PKDD 2012*, pp. 116–131.

Apostolopoulos, Y., Lich, K.H. & Lemke, M.K. (2020). *Complex Systems and Population Health*. Oxford University Press.

Arzate Cruz, C. & Igarashi, T. (2020). A survey on interactive reinforcement learning: Design principles and open challenges. In *Proceedings of the 2020 ACM designing interactive systems conference*, pp. 1195–1209.

Bellman, R. (1957). A Markovian decision process. *J. Math. Mech.*, 679–684.

Chakraborty, B. & Murphy, S.A. (2014). Dynamic treatment regimes. *Annual Rev. Stat. Appl.*, **1**, 447–464.

Chesnaye, N.C., Stel, V.S., Tripepi, G., Dekker, F.W., Fu, E.L., Zoccali, C. & Jager, K.J. (2022). An introduction to inverse probability of treatment weighting in observational research. *Clin. Kidney J.*, **15**(1), 14–20.

Cheung, Y.K., Chakraborty, B. & Davidson, K.W. (2015). Sequential multiple assignment randomized trial (SMART) with adaptive randomization for quality improvement in depression treatment program. *Biometrics*, **71**(2), 450–459.

Clifton, J. & Laber, E. (2020). Q-learning: Theory and applications. *Annual Review of Statistics and Its Application*, **7**, 279–301.

Coronato, A., Naeem, M., De Pietro, G. & Paragliola, G. (2020). Reinforcement learning for intelligent healthcare applications: A survey. *Artif. Intell. Med.*, **109**.

Deliu, N. & Chakraborty, B. 2022. Dynamic treatment regimes for optimizing healthcare. In *The Elements of Joint Learning and Optimization in Operations Management*, Springer, pp. 391–444.

Deliu, N., Williams, J.J. & Chakraborty, B. (2024). Reinforcement learning in modern biostatistics: Constructing optimal adaptive interventions. *Int. Stat. Rev.*

Eckardt, J.-N., Wendt, K., Bornhaeuser, M. & Middeke, J.M. (2021). Reinforcement learning for precision oncology. *Cancers*, **13**(18), 4624.

Ernst, D., Geurts, P. & Wehenkel, L. (2005). Tree-based batch mode reinforcement learning. *J. Machine Learn. Res.*, **6**, 503–556.

Ertefaie, A. & Strawderman, R.L. (2018). Constructing dynamic treatment regimes over indefinite time horizons. *Biometrika*, **105**(4), 963–977.

Fürnkranz, J., Hüllermeier, E., Cheng, W. & Park, S.-H. (2012). Preference-based reinforcement learning: A formal framework and a policy iteration algorithm. *Machine Learn.*, **89**, 123–156.

Fu, Z., Qi, Z., Wang, Z., Yang, Z., Xu, Y. & Kosorok, M.R. 2022. Offline reinforcement learning with instrumental variables in confounded Markov decision processes. https://arxiv.org/abs/2209.08666

Gabriel, I. (2020). Artificial intelligence, values, and alignment. *Minds Mach.*, **30**(3), 411–437.

Garcia, F. & Rachelson, E. (2013). Markov decision processes. *Markov Decis. Processes Artif. Intell.*, 1–38.

Gaweda, A.E., Muezzinoglu, M.K., Aronoff, G.R., Jacobs, A.A., Zurada, J.M. & Brier, M.E. (2005). Incorporating prior knowledge into q-learning for drug delivery individualization. In *Fourth International Conference on Machine Learning and Applications (ICMLA'05)*, pp. 6–pp, IEEE.

Gilliland, C.T., White, J., Gee, B., Kreeftmeijer-Vegter, R., Bietrix, F., Ussi, A.E., Hajduch, M., Kocis, P., Chiba, N. & Hirasawa, R. 2019. The fundamental characteristics of a translational scientist.

Grondman, I., Busoniu, L., Lopes, G.A.D. & Babuska, R. (2012). A survey of actor-critic reinforcement learning: Standard and natural policy gradients. *IEEE Trans. Syst. Man Cybern. C*, **42**(6), 1291–1307.

Hernández-Lerma, O. & Lasserre, J.B. (2012). *Discrete-Time Markov Control Processes: Basic Optimality Criteria*, Vol. 30. Springer Science & Business Media.

Hernan, M.A. & Robins, J.M. (2023). *Causal Inference: What If*, Chapman & Hall/CRC Monographs on Statistics & Applied Probab. CRC Press.

Holzinger, A. (2016). Interactive machine learning for health informatics: When do we need the human-in-the-loop? *Brain Inform.*, **3**(2), 119–131.

Holzinger, A., Langs, G., Denk, H., Zatloukal, K. & Müller, H. (2019). Causability and explainability of artificial intelligence in medicine. *Wiley Interdiscip. Rev. Data Min. Knowl. Discov.*, **9**(4), e1312.

Istepanian, R., Laxminarayan, S. & Pattichis, C.S. (2007). *M-Health: Emerging Mobile Health Systems*. Springer Science & Business Media.

Johnson, AEW, Pollard, T.J., Shen, L., Lehman, L.-H., Feng, M., Ghassemi, M., Moody, B., Szolovits, P., Anthony Celi, L. & Mark, R.G. (2016). MIMIC-III: A freely accessible critical care database. *Sci. Data*, **3**(1), 1–9.

Kahkoska, A.R., Freeman, N.ikkiL.B. & Lich, K.H. (2022). Systems-aligned precision medicine-building an evidence base for individuals within complex systems. In *JAMA health forum*, Vol. 3, American Medical Association.

Kahkoska, A.R., Lich, K.H. & Kosorok, M.R. (2022). Focusing on optimality for the translation of precision medicine. *J. Clin. Trans. Sci.*, **6**(1).

Komorowski, M., Gordon, A., Celi, L.A. & Faisal, A. (2016). A Markov decision process to suggest optimal treatment of severe infections in intensive care. In *Neural information processing systems workshop on machine learning for health*.

Kosorok, M.R. & Laber, E.B. (2019). Precision medicine. *Annual Rev. Stat. Appl.*, **6**, 263–286.

Kosorok, M.R. & Moodie, E.E.M. (2015). *Adaptive Treatment Strategies in Practice: Planning Trials and Analyzing Data for Personalized Medicine*. SIAM.

Laber, E.B., Linn, K.A. & Stefanski, L.A. (2014). Interactive model building for Q-learning. *Biometrika*, **101**(4), 831–847.

Laber, E.B., Lizotte, D.J., Qian, M., Pelham, W.E. & Murphy, S.A. (2014). Dynamic treatment regimes: Technical challenges and applications. *Electron. J. Stat.*, **8**(1), 1225.

Laber, E.B. & Staicu, A.-M. (2017). Functional feature construction for individualized treatment regimes. *J. Amer. Stat. Assoc.*, **113**(523), 1219–1227.

Li, L., Albert-Smet, I. & Faisal, A.A. 2020. Optimizing medical treatment for sepsis in intensive care: From reinforcement learning to pre-trial evaluation.

Li, Z., Chen, J., Laber, E., Liu, F. & Baumgartner, R. (2023). Optimal treatment regimes: A review and empirical comparison. *Int. Stat. Rev.*, **91**(3), 427–463.

Li, G., Gomez, R., Nakamura, K. & He, B. (2019). Human-centered reinforcement learning: A survey. *IEEE Trans. Human-Machine Syst.*, **49**(4), 337–349.

Li, L., Komorowski, M. & Faisal, A.A. 2018. The actor search tree critic (ASTC) for off-policy POMDP learning in medical decision making.

Li, L., Komorowski, M. & Faisal, A.A. 2019. Optimizing sequential medical treatments with auto-encoding heuristic search in POMDPS.

Linn, K.A., Laber, E.B. & Stefanski, L.A. (2015). IQLEARN: Interactive Q-learning in R. *J. Stat. Softw.*, **64**(1).

Lizotte, D.J. & Laber, E.B. (2016). Multi-objective Markov decision processes for data-driven decision support. *J. Mach. Learn. Res.*, **17**(1), 7378–7405.

Love, T., Ajoodha, R. & Rosman, B. (2023). Who should I trust? Cautiously learning with unreliable experts. *Neural Comput. Applicat.*, **35**(23), 16865–16875.

Luckett, D.J., Laber, E.B., Kahkoska, A.R., Maahs, D.M., Mayer-Davis, E. & Kosorok, M.R. (2019). Estimating dynamic treatment regimes in mobile health using V-learning. *J. Amer. Stat. Assoc.*

Luckett, D.J., Laber, E.B., Kim, S. & Kosorok, M.R. (2021). Estimation and optimization of composite outcomes. *J. Mach. Learn. Res.*, **22**(167), 1–40.

Maadi, M., Akbarzadeh Khorshidi, H. & Aickelin, U. (2021). A review on human-AI interaction in machine learning and insights for medical applications. *Int. J. Environ. Res. Public Health*, **18**(4), 2121.

Milani Fard, M. & Pineau, J. (2011). Non-deterministic policies in Markovian decision processes. *J. Artif. Intell. Res.*, **40**, 1–24.

Monahan, G.E. (1982). State of the art-a survey of partially observable Markov decision processes: Theory, models, and algorithms. *Manag. Sci.*, **28**(1), 1–16.

Murphy, S.A. (2003). Optimal dynamic treatment regimes. *J. Royal Stat. Soc. Ser. B: Stat. Methodol.*, **65**(2), 331–355.

Murphy, S.A. (2005a). An experimental design for the development of adaptive treatment strategies. *Stat. Med.*, **24**(10), 1455–1481.

Murphy, S.A. (2005b). A generalization error for Q-learning. *J. Mach. Learn. Res.*, **6**, 1073–1097.

Nahum-Shani, I., Smith, S.N., Spring, B.J., Collins, L.M., Witkiewitz, K., Tewari, A. & Murphy, S.A. (2018). Just-in-time adaptive interventions (JITAIS) in mobile health: Key components and design principles for ongoing health behavior support. *Annals Behav. Med.*, 1–17.

Neuberg, L.G. (2003). Causality: Models, reasoning, and inference, by Judea pearl, Cambridge University Press, 2000. *Econom. Theory*, **19**(4), 675–685.

Nivot, C. (2016). Analyse et étude des processus markoviens décisionnels. Ph.D. Thesis.

Ormoneit, D. & Sen, S. (2002). Kernel-based reinforcement learning. *Machine Learn.*, **49**, 161–178.

Peng, X., Ding, Y., Wihl, D., Gottesman, O., Komorowski, M., Li-wei, H.L., Ross, A., Faisal, A. & Doshi-Velez, F. (2018). Improving sepsis treatment strategies by combining deep and kernel-based reinforcement learning. In *AMIA Annual Symposium Proceedings*, Vol. **2018**, pp. 887, American Medical Informatics Association.

Perera, D., Liu, S., See, K.C. & Feng, M. (2025). Smart imitator: Learning from imperfect clinical decisions. *J. Amer. Med. Inform. Assoc.*, ocae320.

Raghu, A., Komorowski, M., Ahmed, I., Celi, L., Szolovits, P. & Ghassemi, M. 2017. Deep reinforcement learning for sepsis treatment.

Raghu, A., Komorowski, M., Celi, L.A., Szolovits, P. & Ghassemi, M. (2017). Continuous state-space models for optimal sepsis treatment: A deep reinforcement learning approach. In *Machine learning for healthcare conference*, pp. 147–163, PMLR.

Raghu, A., Komorowski, M. & Singh, S. 2018. Model-based reinforcement learning for sepsis treatment. https://arxiv.org/abs/1811.09602

Rehg, J.M., Murphy, S.A. & Kumar, S. (2017). *Mobile health*. Springer International Publishing: Cham.

Robins, J.M. (1986). A new approach to causal inference in mortality studies with a sustained exposure period–application to control of the healthy worker survivor effect. *Math. Modell.*, **7**(9-12), 1393–1512.

Robins, J.M. (1989). The analysis of randomized and non-randomized aids treatment trials using a new approach to causal inference in longitudinal studies. *Health Serv. Res. Methodol. Focus AIDS*, 113–159.

Robins, J.M. (1992). Estimation of the time-dependent accelerated failure time model in the presence of confounding factors. *Biometrika*, **79**(2), 321–334.

Robins, J.M. (1994). Correcting for non-compliance in randomized trials using structural nested mean models. *Commun. Stat. Theory Meth.*, **23**(8), 2379–2412.

Robins, J.M. (1998). Correction for non-compliance in equivalence trials. *Stat. Med.*, **17**(3), 269–302.

Robins, J.M. 2000. Marginal structural models versus structural nested models as tools for causal inference. In *Statistical Models in Epidemiology, the Environment, and Clinical Trials*, Springer, pp. 95–133.

Robins, J.M. (2004). Optimal structural nested models for optimal sequential decisions. In *Proceedings of the second Seattle symposium in biostatistics: Analysis of correlated data*, pp. 189–326, Springer.

Roggeveen, L., El Hassouni, A., Ahrendt, J., Guo, T., Fleuren, L., Thoral, P., Girbes, A.R.J., Hoogendoorn, M. & Elbers, P.W.G. (2021). Transatlantic transferability of a new reinforcement learning model for optimizing hemodynamic treatment for critically ill patients with sepsis. *Artif. Intell. Med.*, **112**, 102003.

Schulte, P.J., Tsiatis, A.A., Laber, E.B. & Davidian, M. (2014). Q- and A-learning methods for estimating optimal dynamic treatment regimes. *Stat. Sci. A Rev. J. Inst. Math. Stat.*, **29**(4), 640.

Shah, S.I.H., Coronato, A. & Naeem, M. (2022). Inverse reinforcement learning based approach for investigating optimal dynamic treatment regime. In *Workshops at 18th international conference on intelligent environments (IE2022)*, pp. 266–276, IOS Press.

Shin, D., Brown, D.S. & Dragan, A.D. 2022. Offline preference-based apprenticeship learning.

Shortreed, S.M., Laber, E., Lizotte, D.J., Stroup, T.S., Pineau, J. & Murphy, S.A. (2011). Informing sequential clinical decision-making through reinforcement learning: An empirical study. *Machine Learn.*, **84**(1-2), 109.

Sonabend, A., Lu, J., Celi, L.A., Cai, T. & Szolovits, P. (2020). Expert-supervised reinforcement learning for offline policy learning and evaluation. *Adv. Neural Inform. Process. Syst.*, **33**, 18967–18977.

Sperger, J., Freeman, N.L.B., Jiang, X., Bang, D., de Marchi, D. & Kosorok, M.R. (2020). The future of precision health is data-driven decision support. *Stat. Anal. Data Mining ASA Data Sci. J.*, **13**(6), 537–543.

Stensrud, M.J., Laurendeau, J.D. & Sarvet, A.L. (2024). Optimal regimes for algorithm-assisted human decision-making. *Biometrika*, **111**(4), 1089–1108.

Sutton, R.S. & Barto, A.G. (2018). *Reinforcement Learning: An Introduction*. MIT press.

Tang, S., Modi, A., Sjoding, M. & Wiens, J. (2020). Clinician-in-the-loop decision making: Reinforcement learning with near-optimal set-valued policies. In *International Conference on Machine Learning*, pp. 9387–9396, PMLR.

Terry, S.F. (2015). Obama's precision medicine initiative. *Genetic Test. Molecular Biomark.*, **19**(3), 113.

Tseng, H.-H., Luo, Y., Cui, S., Chien, J.-T., Ten Haken, R.K. & Naqa, I.E. (2017). Deep reinforcement learning for automated radiation adaptation in lung cancer. *Med. Phys.*, **44**(12), 6690–6705.

Tsiatis, A.A., Davidian, M., Holloway, S.T. & Laber, E.B. (2019). *Dynamic Treatment Regimes: Statistical Methods for Precision Medicine*. Chapman and Hall/CRC.

Uehara, M., Shi, C. & Kallus, N. 2022. A review of off-policy evaluation in reinforcement learning.

Wang, L., Zhang, W., He, X. & Zha, H. (2018). Supervised reinforcement learning with recurrent neural network for dynamic treatment recommendation. In *Proceedings of the 24th ACM SIGKDD International Conference on Knowledge Discovery & Data Mining*, pp. 2447–2456.

Watkins, CJCH & Dayan, P. (1992). Q-learning. *Mach. Learn.*, **8**, 279–292.

Weltz, J., Volfovsky, A. & Laber, E.B. (2022). Reinforcement learning methods in public health. *Clin. Therapeut.*, **44**(1), 139–154.

Wu, X., Li, R., He, Z., Yu, T. & Cheng, C. (2023). A value-based deep reinforcement learning model with human expertise in optimal treatment of sepsis. *Npj Digit. Med.*, **6**(1), 15.

Yang, M.-J., Sutton, S.K., Hernandez, L.M., Jones, S.R., Wetter, D.W., Kumar, S. & Vinci, C. (2023). A just-in-time adaptive intervention (JITAI) for smoking cessation: Feasibility and acceptability findings. *Addict. Behav.*, **136**, 107467.

Yu, C., Dong, Y., Liu, J. & Ren, G. (2019). Incorporating causal factors into reinforcement learning for dynamic treatment regimes in HIV. *BMC Med. Inform. Decision Making*, **19**, 19–29.

Yu, C., Liu, J., Nemati, S. & Yin, G. (2021). Reinforcement learning in healthcare: A survey. *ACM Comput. Surv. (CSUR)*, **55**(1), 1–36.

Yu, C., Ren, G. & Dong, Y. (2020). Supervised-actor-critic reinforcement learning for intelligent mechanical ventilation and sedative dosing in intensive care units. *BMC Med. Inform. Decision Making*, **20**(3), 1–8.

Yu, C., Ren, G. & Liu, J. (2019). Deep inverse reinforcement learning for sepsis treatment. In *2019 IEEE International Conference on Healthcare Informatics (ICHI)*, pp. 1–3.

Zhang, J. (2020). Designing optimal dynamic treatment regimes: A causal reinforcement learning approach. In *International Conference on Machine Learning*, pp. 11012–11022, PMLR.

Zhao, Y., Kosorok, M.R. & Zeng, D. (2009). Reinforcement learning design for cancer clinical trials. *Stat. Med.*, **28**(26), 3294–3315.

Zhao, Y.-Q., Zeng, D., Laber, E.B. & Kosorok, M.R. (2015). New statistical learning methods for estimating optimal dynamic treatment regimes. *J. Amer. Stat. Assoc.*, **110**(510), 583–598.